\documentclass[pre,twocolumn,aps,superscriptaddress,showpacs,floatfix]{revtex4}

\usepackage{amsmath}
\usepackage{latexsym}
\usepackage{float}
\usepackage{amssymb}
\usepackage{graphicx}
\usepackage{textcomp}
\usepackage{hyperref}
\usepackage{graphicx}

\begin{document}
\def\be{\begin{equation}}
\def\ee{\end{equation}}     
\def\bfi{\begin{figure}}
\def\efi{\end{figure}}
\def\bea{\begin{eqnarray}}
\def\eea{\end{eqnarray}}

\newcommand{\vk}{\vec{k}}
\newcommand{\rphi}{\widehat{\varphi}}
\newcommand{\iphi}{\widetilde{\varphi}}
\newcommand{\reta}{\widehat{\eta}}
\newcommand{\ieta}{\widetilde{\eta}}
\newcommand{\vphi}{\boldsymbol{\varphi}}
\newcommand{\eRk}{\emph{R}_{\vec{k},\eta}}
\newcommand{\eIk}{\emph{I}_{\vec{k},\eta}}
\newcommand{\ket}[1]{\vert#1\rangle}
\newcommand{\bra}[1]{\langle#1\vert}
\newcommand{\braket}[2]{\langle #1 \vert #2 \rangle}
\newcommand{\ketbra}[2]{\vert #1 \rangle  \langle #2 \vert}
\newcommand{\up}[1]{^}

\title{Condensation vs Ordering: From the Spherical Models to BEC in the Canonical and Grand Canonical Ensemble}  

\author{A. Crisanti}
\email{andrea.crisanti@uniroma1.it}
\affiliation{Dipartimento di Fisica, Universit\`a di Roma Sapienza, P.le Aldo Moro 2, 00185 Rome, Italy}
\affiliation{Istituto dei Sistemi Complessi - CNR, P.le Aldo Moro 2, 00185 Rome, Italy}

\author{A. Sarracino}
\email{alessandro.sarracino@unicampania.it}
\affiliation{Dipartimento di Ingegneria, Universit\`a della Campania ``Luigi Vanvitelli'', via Roma 29, 81031 Aversa (CE), Italy}

\author{M. Zannetti}
\email{mrc.zannetti@gmail.com}
\affiliation{Dipartimento di Fisica ``E. R. Caianiello'',
Universit\`a di Salerno, via Giovanni Paolo II 132, 84084 Fisciano (SA), Italy,}

\begin{abstract}
In this paper we take a fresh look at the long standing issue of the
nature of macroscopic density fluctuations in the grand canonical
treatment of the Bose-Einstein condensation (BEC).  Exploiting the
close analogy between the spherical and mean-spherical models of
magnetism with the canonical and grand canonical treatment of the
ideal Bose gas, we show that BEC stands for
different phenomena in the two ensembles: an ordering transition of the type
familiar from ferromagnetism in the canonical ensemble and
{\it condensation of fluctuations},
i.e. growth of macroscopic fluctuations in a single degree of freedom,
without ordering, in the grand canonical case. We further clarify that this is a manifestation of nonequivalence of the ensembles, due to the existence of long range correlations in the grand canonical one.
Our results shed new light on the recent
experimental realization of BEC in a photon gas, suggesting that the
observed BEC when prepared under grand canonical conditions
is an instance of condensation of fluctuations.
\end{abstract}

\pacs{05.30.Jp; 05.40.-a; 03.75.Hh; 64.60.-i}

\maketitle

\section{Introduction} 

In general statistical ensembles are constructed
to be equivalent in the thermodynamic limit, but there are exceptions to this
rule.
This paper deals with phenomena arising when this equivalence breaks down. 
Pairs of conjugate ensembles are
obtained by controlling the system either by fixing the value of some
extensive quantity, through appropriate isolating walls, or by putting
it in contact with a reservoir of that same quantity.  A familiar
example,  which will be of central interest in the following,
is that of the canonical and grand canonical pair resulting from fixing either the
density of particles or the chemical potential, while keeping the
system thermalized.  

Basically, equivalence holds in situations
in which correlations are short ranged. Then, the central limit theorem 
guarantees that fluctuations of extensive quantities become negligible in the thermodynamic
limit, so that it doesn't matter whether the system is controlled by enforcing a rigid constraint or through the contact with a reservoir~\cite{Huang,Pathria,Puglisi}.
By the same token lack of equivalence is to be expected when correlations are long
ranged. This is a more rare occurrence, but very
interesting since new physics obtains by switching from one
ensemble to the other within a conjugate pair.  
Best known and
recently much studied is the case of systems with long-range
interactions~\cite{Campa}.

There is one instance of nonequivalence which stands apart: The one which
materializes as an
ideal Bose gas (IBG) is driven through the Bose-Einstein condensation (BEC).
In the canonical ensemble (CE) fluctuations of the condensate behave normally 
while in the grand canonical ensemble (GCE) do persist even in the thermodynamic limit
~\cite{Pathria,Ziff}. Although this is an exact result, it is somewhat puzzling because,
dealing with an ideal gas,
it is not at all obvious where the  long range correlations responsible of the
nonequivalence could come from.
An unbiased attitude ought to
advise to take the facts at face value and to inquiry on possible different mechanisms
underlying BEC in the two ensembles.
Instead, due to a widespread aversion to macroscopic fluctuations of an extensive quantity, which are not suppressed by lowering the temperature, the GCE result has been
variously regarded as unacceptable~\cite{Holthaus},
unphysical~\cite{Fujiwara,Ziff,Stringari,Scully} or even
wrong~\cite{Yukalov} and is commonly referred to as the {\it grand
canonical catastrophe}.  

The need to reconsider afresh this matter has been
prompted by the recent observation of BEC in the lab~\cite{Klaers,Schmitt} in a gas of photons
under grand canonical conditions, which has changed the outlook by
producing evidence 
for the existence of the macroscopic fluctuations of the condensate. 
Therefore, after reckoning with the absence of any catastrophe,
the challenge is to uncover the mechanism responsible of the nonequivalence.
Due to the fundamental character of the question posed, we shall leave the experiment in the background and we shall explore the basic issues in the simplest possible context of the uniform IBG in a box of volume $V$, aiming primarily to outline the conceptual framework
needed to approach this interesting and multifaceted problem. 

\section{The Problem} 

At the 
phenomenological level the mechanism of BEC appears to be the 
same in the CE and in the GCE. 
Denoting by $d$, $d^*$ and  $d_0$ the total density, the density in the excited states and the density
in the ground state, respectively, from the obvious identity $d=d^* +
d_0$ follows the sum rule which must be satisfied by the average quantities
irrespective of the choice of the ensemble
\be
\rho  = \langle d^* \rangle + \langle d_0 \rangle,
\label{sumrule.1}
\ee
where $\rho$ stands for $\langle d \rangle$ and the brackets for the average over
either ensemble. The condensate density  $\langle d_0 \rangle$
is called the BEC order parameter.
Now, for space dimensionality $d > 2$ and in the thermodynamic limit, $ \langle d^* \rangle$ is superiorly bounded by a
finite critical value $\rho_c$~\cite{Huang,Pathria}. Consequently, keeping $T$ fixed and using $\rho$ as control parameter, from Eq.~(\ref{sumrule.1}) immediately follows the density-driven BEC
\be
\langle d_0 \rangle   = \left \{ \begin{array}{ll}
         0, \;\; $for$ \;\; \rho  \leq \rho_c,\\
         \rho - \rho_c, \;\; $for$ \;\;  \rho  > \rho_c,
        \end{array}
        \right .
        \label{dens.1}
        \ee
which, we emphasize, holds irrespective of the ensemble. 
Thus, as far as $\langle d_0 \rangle$ is concerned,
CE and GCE are equivalent. However, a striking difference between the two emerges when the fluctuations of $d_0$ are considered, since in the condensed phase, as previously anticipated, one has~\cite{Ziff}
\be
\left \langle \left (d_0  - \langle d_0 \rangle \right )^2 \right \rangle   = \left \{ \begin{array}{ll}
         0, \;\; $in the CE$,\\
         \langle d_0 \rangle^2 \;\; $in the GCE$,
        \end{array}
        \right .
        \label{dens.2}
        \ee
i.e. normal behavior in the CE and macroscopic fluctuations in the GCE. 
 
The crux of the matter is that at this level of observation no insight can be obtained
as to the why fluctuations ought to behave so differently in the two ensembles.
The point of view that we propose in this paper is that the picture is rationalized by
shifting the description to the finer and underlying level of  the field-theoretic microscopic degrees of freedom, which however are not directly observable. 
In order to clarify the interplay of the different levels of description, in the next 
paragraph we shall exploit the analogy with magnetic systems,
where a quite similar and well understood situation arises.

\section{Spherical and Mean-Spherical Model} 

The IBG in the CE and in the GCE
is well known~\cite{KT,Kac2,Cannon} to be closely related
to the spherical and the mean-spherical models of magnetism.
Let ${\cal H}(\boldsymbol{\varphi}) = \int_V d\vec r \, \varphi(\vec r) \left
(-\frac{1}{2}\nabla^2 \right )\varphi(\vec r)$ be the energy function of a classical scalar paramagnet \cite{Ma} in the volume $V$, where $\boldsymbol{\varphi}$ stands for a configuration  of the local unbounded spin variable $\varphi(\vec r)$.
Due to its bilinear character the above Hamiltonian can be diagonalized by Fourier transform ${\cal H} = \frac{1}{V}\sum_{\vec k} k^2|\varphi_{\vec k}|^2$.
In the spherical model (SM) of Berlin and Kac~\cite{BK} a coupling among the modes is
induced by the imposition of an overall constraint on
the square magnetization
${\cal S}(\boldsymbol{\varphi}) = \int_V d \vec r \,
\varphi^2(\vec r) = \frac{1}{V}\sum_{\vec k} |\varphi_{\vec k}|^2$. Then, in thermal equilibrium the statistical ensemble reads
\be
P_\textrm{SM}(\boldsymbol{\varphi}|\sigma) = \frac{1}{Z_\textrm{SM}}
e^{-\beta {\cal H}(\boldsymbol{\varphi})}
\, \delta \left (\mathit{s}(\boldsymbol{\varphi})-\sigma \right ),
\label{Gauss.4}
\ee
where $Z_\textrm{SM}$ is the partition function,
$\mathit{s}(\boldsymbol{\varphi}) = \frac{1}{V}{\cal S}(\boldsymbol{\varphi})$
the square magnetization density and
$\sigma$ a positive number, which usually is set $\sigma = 1$, but here will be
kept free to vary  as a control parameter. In the mean-spherical model (MSM)~\cite{LW,KT} the constraint is imposed in the mean: An ${\cal S}$-dependent exponential bias is introduced in place of the $\delta$ function
\be
P_\textrm{MSM}(\boldsymbol{\varphi}|\sigma) = \frac{1}{Z_\textrm{MSM}}e^{-\beta [{\cal H}(\boldsymbol{\varphi}) +\frac{\kappa}{2} {\cal S}(\boldsymbol{\varphi})]},
\label{Gauss.3}
\ee
and the intensive parameter $\kappa$ conjugate to ${\cal S}$ must be adjusted so as to satisfy the requirement
\be
\langle \mathit{s}(\boldsymbol{\varphi}) \rangle_\textrm{MSM} = \sigma.
\label{msph.1}
\ee
Although it is the common usage to refer to these  as models, it should be clear from Eqs. (\ref{Gauss.4}) and (\ref{Gauss.3}) that we are dealing with two conjugate ensembles, distinguished by conserving or letting to fluctuate the density $\mathit{s}$. Separating the excitations from the ground state contribution $\mathit{s} = \mathit{s}^*+ \mathit{s}_0$, where $\mathit{s}^* = \frac{1}{V^2}\sum_{\vec k \neq 0} |\varphi_{\vec k}|^2$ and $\mathit{s}_0 = \frac{1}{V^2} \varphi_0^2$,
taking the average and using the constraint
$\langle \mathit{s}  \rangle = \sigma$, independently from the choice of the
model there follows the sum rule analogous to Eq.~(\ref{sumrule.1})
\be
\sigma = \langle \mathit{s}^* \rangle + \langle \mathit{s}_0 \rangle.
\label{sumrule.2}
\ee
Therefore, the variables
$\mathit{s}, \mathit{s}^*, \mathit{s}_0$ and $\sigma$ do correspond to the IBG ones
$d, d^*,d_0$ and $\rho$, with the important difference that in
the present context these are composite variables, built in terms of the microscopic 
set of the magnetization components $[\varphi_{\vec k}]$.
Furthermore, also in this case
for $d > 2$ and in the thermodynamic limit the excitation contribution $\langle \mathit{s}^* \rangle$ is superiorly bounded by a finite critical value $\sigma_c$, see Appendices~\ref{appA} and~\ref{appB} for details.
Hence, keeping $T$ fixed and varying $\sigma$, from Eq.~(\ref{sumrule.2}) there follows 
\be
\langle \mathit{s}_0 \rangle    = \left \{ \begin{array}{ll}
         0, \;\; $for$ \;\; \sigma \leq \sigma_c,\\
         \sigma - \sigma_c, \;\; $for$ \;\; \sigma > \sigma_c,
        \end{array}
        \right .
        \label{BEC.001}
        \ee
showing that $\langle \mathit{s}_0 \rangle$ behaves like the BEC order parameter
and that, as far as $\langle \mathit{s}_0 \rangle$ is concerned, the two models are equivalent. 

\begin{figure}[!tb]
\centering
\includegraphics[width=0.8\columnwidth,clip=true]{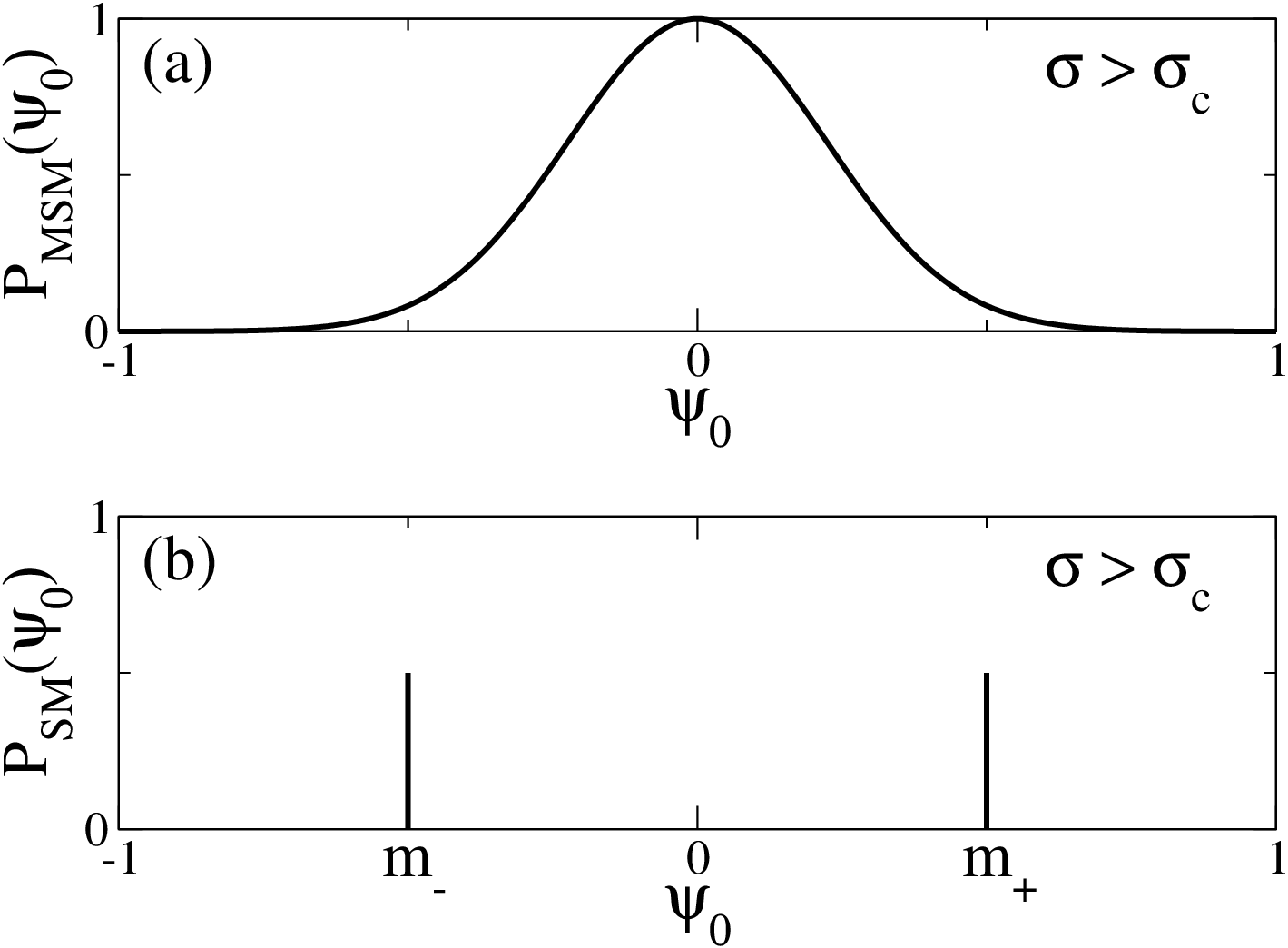}
\caption{Distributions of $\psi_0$ in the MSM model (a) and in the SM model (b)
  for $\sigma > \sigma_c$. The spikes in the bottom panel stand for $\delta$ functions.}
\label{fig1}
\end{figure}

However, at the microscopic level 
a different scenario opens up, since
there is no unique way to form a finite expectation $\langle \mathit{s}_0 \rangle$.
Let us introduce the probability that $\mathit{s}$
takes the value $\sigma^{\prime}$ in the MSM, given by
$K_\textrm{MSM}(\sigma^{\prime}|\sigma) = \int d\boldsymbol{\varphi}
\, P_\textrm{MSM}(\boldsymbol{\varphi}|\sigma)
\delta \left (\mathit{s}(\boldsymbol{\varphi})-\sigma^{\prime} \right )$. Then,
just as a consequence of definitions,
the distributions ~(\ref{Gauss.4}) and~(\ref{Gauss.3}) are related by
\be
P_\textrm{MSM}(\boldsymbol{\varphi}|\sigma) = \int_0^\infty d\sigma^{\prime} \, 
P_\textrm{SM}(\boldsymbol{\varphi}|\sigma^{\prime})
K_\textrm{MSM}(\sigma^{\prime}|\sigma).
\label{kac.1}
\ee
The kernel  has been worked out by
Kac and Thompson~\cite{KT}, obtaining 
$K_\textrm{MSM}(\sigma^{\prime}|\sigma) = \delta(\sigma^{\prime} - \sigma)$ for  $\sigma < \sigma_c$, which implies that the two distributions coincide and, therefore, that the two models are equivalent below $\sigma_c$. Conversely, when $\sigma$ is above $\sigma_c$
the kernel vanishes for $\sigma^{\prime} < \sigma_c$, while for $\sigma^{\prime} > \sigma_c$ is of the spread out form
\be
K_\textrm{MSM}(\sigma^{\prime}|\sigma)    = 
     \frac{e^{-\frac{\sigma^{\prime} - \sigma_c}{2(\sigma - \sigma_c)}}}
          {\sqrt{2\pi (\sigma^{\prime} - \sigma_c)(\sigma - \sigma_c)}}, 
\label{kac.2}
\ee
revealing nonequivalence.
In the following we shall restrict
to the $\sigma > \sigma_c$ domain, where nontrivial behavior is expected.
Integrating out the $\vec k \neq 0$ modes from Eq.~(\ref{kac.1}),
an identical relation between the marginal probabilities of $\psi_0 = \frac{1}{V}\varphi_0$
is obtained. 
In the left hand side there appears  the Gaussian distribution
$P_\textrm{MSM}(\psi_0|\sigma) \propto \exp \{-\beta \kappa V \psi_0^2 /2 \}$,
as it can be verified by inspection from Eq.~(\ref{Gauss.3}),
since $P_\textrm{MSM}(\boldsymbol{\varphi}|\sigma)$
factorizes in Fourier space.
From this follows $\langle \mathit{s}_0 \rangle = (\beta \kappa V)^{-1}$.
So, from the second line of  Eq.~(\ref{BEC.001}) we get
\be
\kappa    = 1/[\beta V (\sigma - \sigma_c)],
\label{kappa.001}
\ee
which implies 
  \be
P_\textrm{MSM}(\psi_0|\sigma) =
         \frac{e^{-\frac{1}{2(\sigma-\sigma_c)}\psi_0^2}}{\sqrt{2\pi (\sigma-\sigma_c)}}.
\label{ensmbl.001}
\ee
Hence, plugging in the explicit expression of $K_\textrm{MSM}(\sigma^{\prime}|\sigma)$
it is not difficult to verify that Eq.~(\ref{kac.1}) is satisfied by the ansatz
\be
P_\textrm{SM}(\psi_0|\sigma) = 
\frac{1}{2}[\delta (\psi_0 - m_-) + 
         \delta (\psi_0 - m_+)],
\label{ensmbl.010}
        \ee
where $m_\pm = \pm \sqrt{\sigma - \sigma_c}$
is the spontaneous magnetization density which would be obtained, for instance, by switching off an external magnetic field~\cite{BK,KT}.
Thus, we have two quite different 
distributions, as it is clearly illustrated by the plots in Fig.~\ref{fig1}. 

We are now in the position to draw some conclusions. The BEC-like order parameter 
$\langle \mathit{s}_0 \rangle$ can be computed microscopically as
the average composite variable $\langle \psi_0^2 \rangle$. Then,
it is straightforward to check from
Eqs.~(\ref{ensmbl.001}) and~(\ref{ensmbl.010}) that 
Eq.~(\ref{BEC.001}) is satisfied in both cases. However, it is enough to take a look at
Fig.~\ref{fig1} to realize that the numerically identical result 
$\langle \psi_0^2 \rangle = (\sigma - \sigma_c)$ for $\sigma > \sigma_c$ in the two models
stands for two different phenomena.   The double peaked distribution of the SM case
is the familiar one for a ferromagnet in the magnetized phase,
each peak being associated to a pure state and with the up-down symmetry of the model
spontaneously broken. Namely, the distribution is the even mixture of these two
pure states.  This means that in the SM the BEC-like transition observed at the level of 
$\langle \mathit{s}_0 \rangle$ is the manifestation of an underlying {\it ordering} transition, 
and that the BEC order parameter is 
the square of the spontaneous magnetization, i.e.
$\langle \psi_0^2 \rangle = m_\pm^2$. By contrast, 
in the MSM case we have the opposite situation, since $\langle \psi_0^2 \rangle$ is the variance
of a broad Gaussian distribution centered on the origin. Therefore there is no ordering, no breaking of the symmetry. In this case the BEC-like transition undergone by  $\langle \mathit{s}_0 \rangle$     is the manifestation of the microscopic
variable $\psi_0$ developing finite fluctuations. 
The reason for this can be grasped intuitively.
In the SM, due to the sharp constraint, there
is enough nonlinearity to produce ordering.
In the MSM framework this cannot be achieved, since the statistics are Gaussian. Then,
the only mean to build up the finite value of $\langle \mathit{s}_0 \rangle$
needed to saturate the sum rule~(\ref{sumrule.2}) above $\sigma_c$ 
is by growing fluctuations in the {\it single} degree of freedom $\psi_0$. Elsewhere~\cite{EPL,CCZ,Zannetti,Merhav,Marsili}, this type of transition, characterized by the fluctuations of an extensive quantity condensing into {\it one} microscopic component, has been referred to as condensation of fluctuations.
The phenomenological picture is completed
by the fluctuations of $\mathit{s}_0$ itself which, as it follows easily from Eqs.~(\ref{ensmbl.001}) and~(\ref{ensmbl.010}), are given by
\be
\left \langle \left (\mathit{s}_0  - \langle \mathit{s}_0 \rangle \right )^2 \right \rangle   = \left \{ \begin{array}{ll}
         0, \;\; $in the SM$,\\
        2 \langle \mathit{s}_0 \rangle^2, \;\; $in the MSM$.
        \end{array}
        \right .
        \label{flcts.2}
        \ee
 Comparing this with Eq.~(\ref{dens.2}), the analogy is evident. However, now no catastrophical behavior can be envisaged, 
 because the fluctuations of $\mathit{s}_0$ are trivially a consequence of the different microscopic statistics
 in the two models. 
 
 Having analysed how the nonequivalence unfolds, the remaining task is to clarify where it does to originate from, which ultimately must be in the presence of long range correlations. 
The explanation is that in the MSM the parameter $\kappa$ is related to the correlation length $\xi$
by $\kappa = \xi^{-2}$~\cite{Ma} and from Eq.~(\ref{kappa.001}) we see that
in the thermodynamic limit
 $\kappa$ vanishes like $1/V$ when $\sigma$ is fixed above $\sigma_c$. 
Therefore, in the entire condensed phase the MSM is critical, while the SM is not.
Hence, the lack of equivalence. We emphasize that the onset of these critical correlations 
in the MSM is the unifying thread behind the BEC-like transition accompanied by macroscopic fluctuations 
of $\mathit{s}_0$.

\section{Back to the IBG} 

We may now go back to the main topic of the IBG
with the advantage of hindsight, since we know what to look for: The microscopic 
variables underlying the phenomenological level, 
in terms of which we expect to expose both the different mechanisms of BEC
in the CE and GCE and the nonequivalence cause. This is accomplished by introducing the creation and destruction operators and by using the representation of the density matrix in the associated coherent state basis.
Let us first diagonalize the energy and number operators by Fourier transform ${\cal H} = \sum_{\vec k}  \epsilon_{\vec k} a_{\vec k}^\dagger a_{\vec k}$ and ${\cal N} = \sum_{\vec k}  a_{\vec k}^\dagger a_{\vec k}$, where $\epsilon_{\vec k}$ is
the single particle energy. In the 
Glauber-Sudarshan P-representation~\cite{Glauber,Sudarshan} the density matrix is given by $D(\rho) = \int d^2\boldsymbol{\alpha} \, P(\boldsymbol{\alpha}|\rho)
\ketbra{\boldsymbol{\alpha}}{\boldsymbol{\alpha}}$, 
where $\ket{\boldsymbol{\alpha}}$
are product states 
$\prod_{\vec k} \ket{\alpha_{\vec k}}$ and the $\vec k$-mode factor $\ket{\alpha_{\vec k}}$ is the eigenvector of the annihilation operator $a_{\vec k}\ket{\alpha_{\vec k}}=\alpha_{\vec k} \ket{\alpha_{\vec k}}$ with complex eigenvalue $\alpha_{\vec k}$.
Then, in the GCE the weight function reads~\cite{Glauber}
\be
 P_\textrm{GCE}(\boldsymbol{\alpha}|\rho) = 
 \prod_{\vec k} \frac{1}{\pi \langle n_{\vec k} \rangle} \exp \left \{-\frac{|\alpha_{\vec k}|^2}
 {\langle n_{\vec k} \rangle} \right \},
 \label{PGCE.1}
 \ee
 where $\langle n_{\vec k} \rangle = [e^{\beta(\epsilon_k - \mu)}-1]^{-1}$ is the usual Bose average occupation number of the state $\ket{\vec k}$~\cite{Huang} and $\mu$ stands for
 the chemical potential.
 Using the identity
 $\langle |\alpha_{\vec k}|^2 \rangle = \langle n_{\vec k} \rangle$, which easily follows
 from  Eq.~(\ref{PGCE.1}), the equation fixing $\mu$ for the given value of $\rho$ 
 reads $\frac{1}{V}\sum_{\vec k} \langle |\alpha_{\vec k}|^2 \rangle
 = \rho$. Since this is nothing but Eq.~(\ref{sumrule.1}), we may 
 write $d = \frac{1}{V}\sum_{\vec k}  |\alpha_{\vec k}|^2$ and, consequently,
 $d_0 = |\eta_0|^2$, after setting $\eta_0 = \frac{1}{\sqrt{V}}\alpha_0$.
 This allows to identify $[\alpha_{\vec k}]$ with the sought for set of microscopic variables
 analogous to $[\varphi_{\vec k}]$.
 Following the magnetic example, we must focus on the statistics of 
 the zero component, keeping in mind however that now is a complex quantity 
 $\eta_0 = |\eta_0|e^{i\theta}$.
 The starting point is the relation between ensembles analogous 
 to Eq.~(\ref{kac.1}) (see Appendices~\ref{appC} and~\ref{appD})
\be
P_\textrm{GCE}(\boldsymbol{\alpha}|\rho) = \int_0^\infty d\rho^{\prime} \, 
P_\textrm{CE}(\boldsymbol{\alpha}|\rho^{\prime})
K_\textrm{GCE}(\rho^{\prime}|\rho),
\label{kac.10}
\ee 
where 
$K_\textrm{GCE}(\rho^\prime|\rho)$ is the probability in the CGE
that the density takes the value $\rho^\prime$. This is known as the Kac function~\cite{Ziff}, whose
form is similar to that of Eq.~(\ref{kac.2}). Namely,
$K_\textrm{GCE}(\rho^{\prime}|\rho) = \delta(\rho^{\prime} - \rho)$ for  $\rho < \rho_c$, while when $\rho$ is above $\rho_c$ it vanishes  
for $\rho^{\prime} < \rho_c$ and is given by
\be
K_\textrm{GCE}(\rho^{\prime}|\rho)    = 
     \frac{e^{-\frac{\rho^{\prime} - \rho_c}{\rho - \rho_c}}}
          {\rho - \rho_c}, \quad \text{for} \quad \rho^{\prime} > \rho_c.
          \label{kac.5}
\ee
The relation between the $\eta_0$ marginal distributions
is then obtained by integrating out  
the $\vec k \neq 0$ modes $P_\textrm{GCE}(\eta_0|\rho) = \int_0^\infty d\rho^{\prime} \, 
P_\textrm{CE}(\eta_0|\rho^{\prime}) \,
K_\textrm{GCE}(\rho^{\prime}|\rho)$. Inserting in the left hand side the contribution 
from the first factor of Eq.~(\ref{PGCE.1})
 \be
 P_\textrm{GCE}(\eta_0|\rho) = 
 \frac{V}{\pi \langle n_{0} \rangle} \exp \left \{-\frac{V|\eta_0|^2}
 {\langle n_{0} \rangle} \right \},
 \label{PGCE.2}
 \ee
 and substituting for $K_\textrm{GCE}(\rho^{\prime}|\rho)$ the above expression, the equation
 is solved by the ansatz
\be
 P_\textrm{CE}(\eta_0|\rho) = 
         \frac{1}{\pi} \, \delta (|\eta_0|^2 - (\rho - \rho_c)).
\label{CE.010}
        \ee
\begin{figure}[!tb]
\centering
\includegraphics[width=0.8\columnwidth,clip=true]{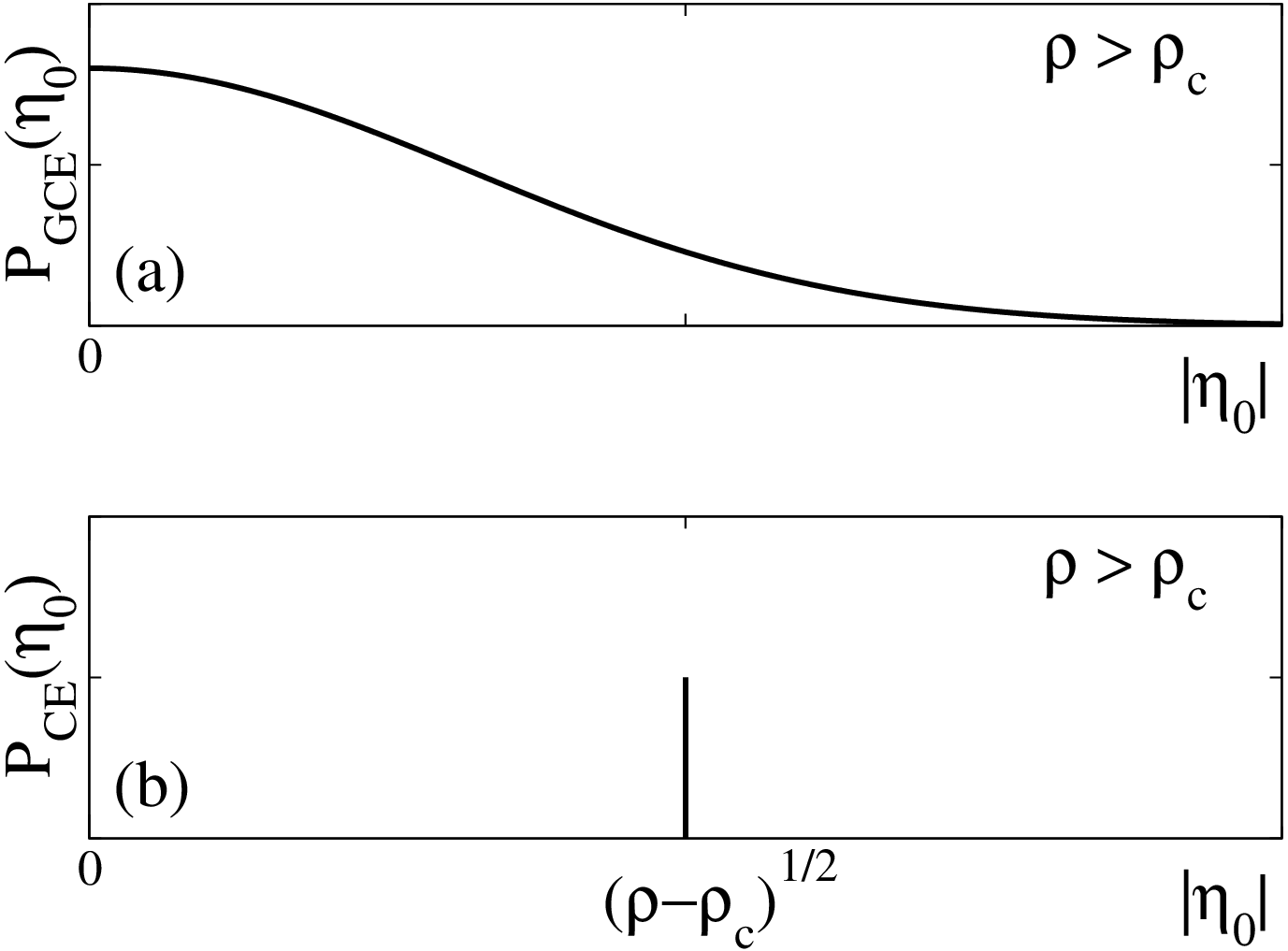}
\caption{Distributions of $|\eta_0|$ in the GCE (a) and in the CE (b)
  for $\rho > \rho_c$. The spike in the bottom panel stands for the $\delta$
  function distribution.}
\label{fig2}
\end{figure}        
From the plot of the two distributions~(\ref{PGCE.2}) and~(\ref{CE.010}) in Fig.~\ref{fig2}, 
we see
by inspection that we are confronted with a situation qualitatively 
similar to the one in the magnetic case.
In both ensembles $|\eta_0|$ develops a nonobservable finite expectation value
\be
\langle |\eta_0| \rangle   = \left \{ \begin{array}{ll}
         \sqrt{\rho - \rho_c}, \;\; $CE$,\\
        \frac{1}{2}\sqrt{\pi(\rho - \rho_c)}, \;\; $GCE$.
        \end{array}
        \right .
        \label{A.2}
        \ee
The observable     
BEC order parameter exhibits the same numerical value as 
in Eq.~(\ref{dens.1})      
\be
\langle d_0 \rangle = \langle |\eta_0|^2 \rangle   = \left \{ \begin{array}{ll}
         (\rho - \rho_c), \;\; $CE$,\\
        (\rho - \rho_c), \;\; $GCE$,
        \end{array}
        \right .
        \label{A.2b}
        \ee
which is achieved through fluctuations in the GCE and without fluctuations in
the CE 
\be        
\left \langle \left (|\eta_0|  - \langle |\eta_0| \rangle \right )^2 \right \rangle   = \left \{ \begin{array}{ll}
         0, \;\; $CE$, \\
        (1-\pi/4)(\rho - \rho_c), \;\; $GCE$.
        \end{array}
        \right .
        \label{A.3}
        \ee
This means that the sum rule~(\ref{sumrule.1}) in the CE is saturated by fixing the modulus 
to the precise finite value
$|\eta_0| = \sqrt{\rho - \rho_c}$. Because of 
this freezing of $|\eta_0|$, BEC in the CE fits into the scheme of an ordering
transition akin to the ferromagnetic transition in the SM. 
Conversely, BEC in the GCE does not take place through ordering. 
Rather,  the saturation of the sum rule 
is achieved  by growing the macroscopic fluctuations of $|\eta_0|$, as Eq.~(\ref{A.3}) shows. 
Therefore, in this case BEC fits into the scheme of the condensation
transition. Ordering is ruled out because the width of the probability distribution persists in the thermodynamic limit.  Moreover, assuming
$\epsilon_k \sim k^{\alpha}$, where the power $\alpha$ depends on the dispersion relation,
for small $k$ and small $\mu$ we may approximate
$\langle n_{\vec k} \rangle^{-1} \simeq \beta [ k^{\alpha}  - \mu]$ and inserting this into 
Eq.~(\ref{PGCE.1}) we have that the chemical potential, like $\kappa$ in the preceding
case, is connected to the correlation length
by $-\mu=\xi^{-\alpha}$. Since the formation of the condensed phase in the $\textrm{GCE}$ requires $\mu$ 
to vanish in the 
thermodynamic limit~\cite{Huang}, we have that 
the condensed phase is critical throughout in the GCE
but not in the CE. This explains
the origin of nonequivalence which, as in the magnetic case, 
is not revealed by the BEC order parameter
but emerges only at the level of the higher cumulant
\be
\left \langle \left (|\eta_0|^2  - \langle |\eta_0|^2 \rangle \right )^2 \right \rangle   = \left \{ \begin{array}{ll}
        (\rho - \rho_c)^2, \;\; $GCE$,\\
        0, \;\; $CE$.
        \end{array}
        \right .
        \label{A.4}
        \ee     
Hence, the phenomenological result of Eq.~(\ref{dens.2}), rather than being pathological,
is now accounted for as a byproduct of the critical correlations in the GCE.

\section{Summary} 

In this paper we have investigated the differences arising
when BEC in a homogeneous IBG is treated in the CE and in the GCE. The analysis has been
carried out by taking advantage of the close analogy with the
the spherical and mean spherical models of magnetism.
The problem is of particular interest because the ensemble nonequivalence issue encroaches
the fundamental question of the nature of BEC. We have shown that
ordering takes place in the CE, while condensation takes place in the GCE, whose prominent manifestation are the macroscopic
fluctuations of the condensate.
Therefore, we suggest that the recent experimental realization of BEC in a gas of photons \cite{Klaers,Schmitt,Klaers2}
ought to be regarded as qualitatively different from other experimental instances of BEC,
such as those with cold atoms, precisely because the grand canonical conditions
lead to BEC as condensation of fluctuations. 
Moreover, by retracing the origin of nonequivalence to the onset of critical correlations
in the condensed phase of the GCE, we have pointed out that the observable phenomenology follows as a consequence. So, knowledge of the existence of these 
correlations could possibly serve as a useful guide in the planning of future experiments.
As a final remark, notice that the above analysis has involved the modulus, but not the phase of $\eta_0$.
 This means that the distinction between ordering and condensation is decoupled
 from the issue of the breaking of the gauge symmetry. This is a separate and important
 problem which will be the object of future work.

\acknowledgements{Informative and quite useful conversations on photons BEC  with Prof. Claudio Conti are gratefully acknowledged. AS acknowledges
support from the ``Programma VALERE'' of University of Campania
``L. Vanvitelli'' and from MIUR PRIN project ``Coarse-grained description for non-equilibrium systems and transport phenomena (CO-NEST)'' n. 201798CZLJ.}

\appendix

\section{Spherical and Mean Spherical Model}
\label{appA}

Let  ${\boldsymbol \varphi}$ be a configuration of the magnetization field
$\varphi(\vec r)\in (-\infty,+\infty)$ over an hypercube
${\vec r} \in V \subset \mathbb{R}^d$ of side $L$, whose energy is given by
\begin{equation}
{\cal H}(\boldsymbol{\varphi}) = \int_V d\vec r \,
\varphi(\vec r) \left (-\frac{1}{2}\nabla^2 \right )\varphi(\vec r).
\label{H0}
\end{equation}
Ensembles, or models, are defined by thermalizing the system and specifying conditions
imposed on the overall square magnetization 
\be
   {\cal S}(\boldsymbol{\varphi}) = \int_V d{\vec r}\, \varphi^2({\vec r}).
   \label{sqmagn.1}
   \ee
   The spherical model (SM) of Berlin and Kac~\cite{BK}, which corresponds to the ensemble canonical with respect to ${\cal S}(\boldsymbol{\varphi})$,
   is obtained by imposing a sharp constraint on the sqaure magnetization density
   \be
   \mathit{s}(\boldsymbol{\varphi}) = \sigma,
   \label{sphconstr.1}
   \ee
   where $\mathit{s}(\boldsymbol{\varphi}) = \frac{1}{V}{\cal S}(\boldsymbol{\varphi})$ and
      $\sigma$ is a positive number. This leads to the probability distribution
\be
 P_{\textrm{SM}}({\boldsymbol \varphi}) = \frac{1}{Z_{\textrm{SM}}}\, e^{-\beta {\cal H}({\boldsymbol \varphi})}\,
 \delta \left(\mathit{s}(\boldsymbol{\varphi})- \sigma\right),
 \label{spher.1}
 \ee
with the partition function
\be
 Z_{\textrm{SM}}=  \int {\cal D} {\boldsymbol \varphi}\, e^{-\beta {\cal H}({\boldsymbol \varphi})}\,
                          \delta \left( \mathit{s}(\boldsymbol{\varphi}) - \sigma\right).
                          \label{smpartfct.1}
                          \ee
The mean spherical model (MSM) of Lewis and Wannier~\cite{LW}, which corresponds to the ensemble grand canonical with respect to ${\cal S}(\boldsymbol{\varphi})$,
is defined by
\be
P_{\textrm{MSM}}({\boldsymbol \varphi}|\sigma) = \frac{1}{Z_{\textrm{MSM}}}\,
\exp\left\{-\beta [{\cal H}({\boldsymbol \varphi})
                          +\frac{\kappa}{2}\,{\cal S}(\boldsymbol{\varphi})]\right\},
                          \label{GCE.1}
\ee
with the partition function
\be
 Z_{\textrm{MSM}} =  \int {\cal D} {\boldsymbol \varphi}\, \exp\left\{-\beta [{\cal H}({\boldsymbol \varphi})
                         +\frac{\kappa}{2}\,{\cal S}(\boldsymbol{\varphi})]\right\},
                         \label{ZG}
\ee
and where the parameter $\kappa$ is determined self-consistently by imposing the 
constraint~(\ref{sphconstr.1}) on average
\be
    \langle \mathit{s}(\boldsymbol{\varphi}) \rangle_{\textrm{MSM}} = \sigma,
   \label{sph_con}
   \ee
   as it is explained in the next section. Notice that in both models $\beta$ and $\sigma$
   are control parameters.

\section{Solution of the mean spherical model}
\label{appB}

Imposing
periodic boundary conditions and postulating the existence of a
microscopic length $a_0$, the allowed wave vectors of the Fourier
components
\be
\varphi_{\vec k} = \int_V d\vec r \, \varphi(\vec r) \, e^{i\vec k \cdot \vec r}
\label{F.1}
\ee
are given by
\be
\vec k = \frac {2\pi}{L} \vec n, \;\;\; n_i=0,\pm 1, \pm 2,..., \pm N_{\rm max},
\label{wvect.1}
\ee
where $N_{\rm max} = L/a_0$ supposedly is an integer. The inverse transform reads
\be
\varphi(\vec r) = \frac{1}{V} \sum_{\vec k} \varphi_{\vec k} \, e^{-i\vec k \cdot \vec r}.
\label{F.2}
\ee
Using
\be
\delta_{\vec k,\vec {k}^{\prime}} = \frac{1}{V} \int_V d\vec r \,  e^{i(\vec k -\vec {k}^{\prime})\cdot \vec r},
\label{F.3}
\ee
and
\be
\delta(\vec r - \vec{r}^{\prime}) = \frac{1}{V} \sum_{\vec k}  e^{-i\vec k \cdot (\vec r- \vec{r}^{\prime})},
\label{F.4}
\ee
the energy function~(\ref{H0})  and  ${\cal S}(\boldsymbol{\varphi})$  take diagonal forms
\be
{\cal H}(\boldsymbol{\varphi}) = \frac{1}{2V}\sum_{\vec k} k^2 |\varphi_{\vec k}|^2,
\;\; {\cal S}(\boldsymbol{\varphi}) = \frac{1}{V}\sum_{\vec k} |\varphi_{\vec k}|^2,
\label{F.5}
\ee
where we have used the reality of $\varphi(\vec r)$ and Eq.~(\ref{F.1}), which imply
$\varphi_{-\vec k} = \varphi^*_{\vec k}$.

The partition function of the MSM can be computed straightforwardly from Eq.~(\ref{ZG})
\be
\label{eq:ZGe}
  Z_\textrm{MSM} = (2\pi V)^{N/2} \prod_{\vec k} \frac{1}{\sqrt{\beta (k^2 + \kappa)}}, 
  \qquad {\rm Re}\ \kappa > 0,
\ee
where $N=\sum_{\vec k} 1$ is the total number of modes.
Using this expression and
\be
\label{eq:sp_2}
\langle \mathit{s}(\boldsymbol{\varphi})\rangle_\textrm{MSM} =
-\frac{2}{\beta V}\frac{\partial}{\partial \kappa}\, \ln Z_\textrm{MSM},
\ee
the mean spherical constraint~(\ref{sph_con}) reads   
\be
  \beta \sigma = \frac{1}{V}\sum_{\vec k} \frac{1}{(k^2 + \kappa)}
               = \frac{1}{V\kappa} + \frac{1}{V}\sum_{\vec k\not= 0} \frac{1}{k^2 + \kappa},
               \label{split.1}
\ee
where we have separated the $\vec k = 0$ contribution from the rest.
Then, for $V$ sufficiently large, the sum over $\vec k \neq 0$ can be replaced by an integral
\be
\frac{1}{V}\sum_{\vec k\not= 0} \frac{1}{k^2 + \kappa} \rightarrow  B(\kappa) = \int_{0}^{\Lambda}  \frac{d\mu(k)}{(k^2 + \kappa)},
\ee
and Eq.~(\ref{split.1}) can be recast as
\be
\label{eq:Sp_C}
  \beta \sigma = \frac{1} {V\kappa} +  B(\kappa),
\ee
with the integration measure defined by $d\mu(k) = \Omega_d k^{d-1}$, where $\Omega_d=[2^{d-1}\pi^{d/2}\Gamma(d/2)]$ is the $d$-dimensional solid angle
and $\Lambda \sim 1/a_0$ is a cutoff.
The function $B(\kappa)$ is a monotonic decreasing function of $\kappa$, which diverges at
$\kappa =0$ for $d \leq 2$, while its
 maximum value at $\kappa =0$ for $d > 2$ is given by
\be
  B(0) =  \int_{0}^{\Lambda}  \frac{d\mu(k)}{k^2}
 = \Omega_d \Lambda^{d-2}/(d-2).
    \label{B.1}
    \ee
    Therefore, for $d \leq 2$ the first term in the right hand side of Eq.~(\ref{eq:Sp_C})
    can be neglected for any choice of $\beta$ and $\sigma$, and the solution is given by
    \be
    \widetilde{\kappa} = B^{-1}(\beta \sigma),
    \label{critical.01}
    \ee
    where $B^{-1}$ is the inverse function of $B(\kappa)$. Instead, for $d>2$,
    the condition $\beta\sigma=B(0)$ defines a
    critical line on the $(\beta,\sigma)$ plane below which the solution is still
    given by Eq.~(\ref{critical.01}), while above it is necessary to retain also the
    first term in the right hand side of Eq.~(\ref{eq:Sp_C}). 
    Thus, keeping $\beta$ fixed, the critical value of $\sigma$ is given by
    \be
    \sigma_c=\beta^{-1}B(0),
    \label{critical.1}
    \ee
    and the full solution of Eq.~(\ref{eq:Sp_C}) reads
 \be
  \widetilde{\kappa} 
    = \begin{cases}
      B^{-1}(\beta \sigma), \,\, $for$ \,\,  \sigma < \sigma_c,\\
      (B_1V)^{-2/d}, \,\, $for$ \,\,  \sigma = \sigma_c,\\
      1/[\beta V (\sigma - \sigma_c)], \,\, $for$ \,\, \sigma > \sigma_c,
    \end{cases}
    \label{B.2}
    \ee   
    where $B_1$ is a positive constant.
\begin{figure}[!tb]
\centering
\includegraphics[width=0.8\columnwidth,clip=true]{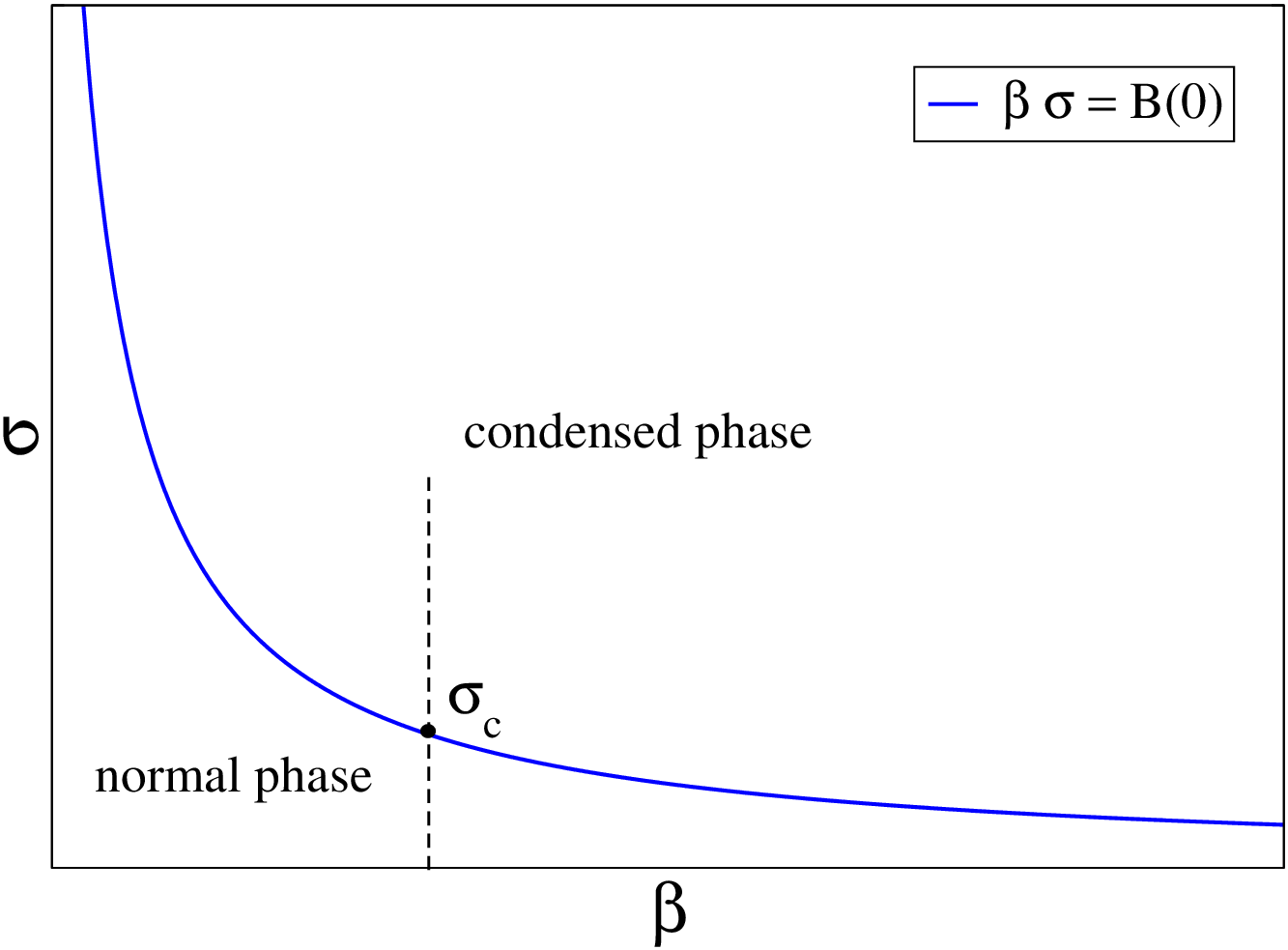}
\caption{Phase diagram for $d > 2$. The dashed vertical line shows
  the thermodynamic path of the transition driven by $\sigma$ while keeping
$\beta$ fixed.}
\label{figSM}
\end{figure}

As a consequence of the mode independence, implied by Eq. (\ref{F.5}), the 
$P_\textrm{MSM}({\boldsymbol \varphi}|\sigma)$ distribution factorizes. Therefore,
introducing the notation $\psi_0 = \frac{1}{V}\varphi_0$ for the magnetization density, the 
$\vec k = 0$ contribution is given by
\be
P_\textrm{MSM}({\psi_0}|\sigma) = \sqrt{\frac{\beta \widetilde{\kappa}V}{2\pi}}
e^{-\frac{\beta \widetilde{\kappa}V}{2}\psi^2_0},
\label{distr.1}
\ee
and inserting the result (\ref{B.2}) for $\widetilde{\kappa}$, in the $V \to \infty$ limit the result reported in the main text is obtained
\be
\lim_{V \to \infty}P_\textrm{MSM}(\psi_0|\sigma) = \left \{ \begin{array}{ll}
         \delta(\psi_0), \;\; $for$ \;\; \sigma \leq \sigma_c,\\
         \frac{e^{-\frac{1}{2(\sigma-\sigma_c)}\psi_0^2}}{\sqrt{2\pi (\sigma-\sigma_c)}}, 
         \;\; $for$ \;\; \sigma > \sigma_c.
        \end{array}
        \right .
        \label{ensmbl.001bis}
        \ee

\section{The SM - MSM connection}
\label{appC}

Using the identity $\int_0^{\infty} d \sigma^{\prime} \, \delta (\mathit{s}({\boldsymbol \varphi})
- \sigma^{\prime} ) = 1$, the MSM ensemble~(\ref{GCE.1}) can be
rewritten as 
\begin{eqnarray}
&&P_{\textrm{MSM}}({\boldsymbol \varphi}|\sigma) \nonumber \\
 & = & \int_0^{\infty} d \sigma^{\prime} \,
\frac{1}{Z_{\textrm{MSM}}(\sigma)} \,
\exp\left\{-\beta [{\cal H}({\boldsymbol \varphi})
                          +\frac{\widetilde{\kappa}(\sigma)}{2}\,{\cal S}(\boldsymbol{\varphi})]\right\} \nonumber \\
                          &\times&\delta (\mathit{s}({\boldsymbol \varphi}) - \sigma^{\prime}) \nonumber \\
&=&  \int_0^{\infty} d \sigma^{\prime} \, \left [ \frac{1}{Z_\textrm{SM}(\sigma^{\prime})} \,
e^{-\beta {\cal H}({\boldsymbol \varphi})} \delta (\mathit{s}({\boldsymbol \varphi}) - \sigma^{\prime}) \right ]   \nonumber \\
&\times& \, \left [e^{-\beta \frac{\widetilde{\kappa}(\sigma)}{2} V  \sigma^{\prime}}
\frac{ Z_\textrm{SM}(\sigma^{\prime})}{ Z_\textrm{MSM}(\sigma)} \right ] ,          
                          \label{connection.1}
                          \end{eqnarray}
which, using the definition~(\ref{smpartfct.1}) of the SM partition function,
can be further manipulated as
\begin{eqnarray}
P_{\textrm{MSM}}({\boldsymbol \varphi}|\sigma) & = &
 \int_0^{\infty} d \sigma^{\prime} \, \left [ \frac{1}{Z_\textrm{SM}(\sigma^{\prime})} \,
e^{-\beta {\cal H}({\boldsymbol \varphi})} \delta (\mathit{s}({\boldsymbol \varphi}) - \sigma^{\prime}) \right ] \, \nonumber \\
&\times&\int {\cal D} {\boldsymbol \varphi^{\prime} }\, \frac{e^{-\beta \left [{\cal H}({\boldsymbol \varphi^{\prime} }) + \frac{\widetilde{\kappa}(\sigma)}{2} V  \sigma^{\prime} \right ] }}{{ Z_\textrm{MSM}(\sigma)}}
\, \delta \left( \mathit{s}(\boldsymbol{\varphi^{\prime} }) - \sigma^{\prime} \right) \nonumber \\
& = & 
 \int_0^{\infty} d \sigma^{\prime} \, \left [ \frac{1}{Z_\textrm{SM}(\sigma^{\prime})} \,
e^{-\beta {\cal H}({\boldsymbol \varphi})} \delta (\mathit{s}({\boldsymbol \varphi}) - \sigma^{\prime}) \right ] \, \nonumber \\
&\times& \int {\cal D} {\boldsymbol \varphi^{\prime} }\, \frac{e^{-\beta \left [{\cal H}({\boldsymbol \varphi^{\prime} }) + \frac{\widetilde{\kappa}(\sigma)}{2} {\cal S}(\boldsymbol{\varphi^{\prime}}) \right ] }}{{ Z_\textrm{MSM}(\sigma)}}
\, \nonumber \\
&\times& \delta \left( \mathit{s}(\boldsymbol{\varphi^{\prime} }) - \sigma^{\prime} \right). \label{connection.2}
                          \end{eqnarray}   
Recognizing that in the square bracket there appears $P_{\textrm{SM}}({\boldsymbol \varphi}|\sigma^{\prime})$, while the last integral 
\be
K_\textrm{MSM}(\sigma'|\sigma) = \int {\cal D} {\boldsymbol \varphi^{\prime} }\, \frac{e^{-\beta \left [{\cal H}({\boldsymbol \varphi^{\prime} }) + \frac{\widetilde{\kappa}(\sigma)}{2} {\cal S}(\boldsymbol{\varphi^{\prime}}) \right ] }}{{ Z_\textrm{MSM}(\sigma)}}
\, \delta \left( \mathit{s}(\boldsymbol{\varphi^{\prime} }) - \sigma^{\prime} \right)
\label{kernl.1}
\ee
is the probability that $\mathit{s}(\boldsymbol{\varphi})$ takes the value $\sigma^{\prime}$
in the MSM parametrized by $\sigma$, the probabilities of
$\boldsymbol{\varphi}$ in the two models are related by
\be
P_\textrm{MSM}({\boldsymbol \varphi}|\sigma) =
\int_{0}^{\infty} d\sigma^{\prime} \, P_\textrm{SM}({\boldsymbol \varphi}|\sigma^{\prime})\,
K_\textrm{MSM}(\sigma^{\prime}|\sigma).
\label{MCE.1}
  \ee
Eliminating the $\varphi_{\vec k \neq 0}$ components from the above equation by integration, a similar relation is obtained between the $\psi_0$ marginal distributions
\be
P_\textrm{MSM}({\psi_0}|\sigma) =
\int_{0}^{\infty} d\sigma^{\prime} \, P_\textrm{SM}({\psi_0}|\sigma^{\prime})\,
K_\textrm{MSM}(\sigma^{\prime}|\sigma).
\label{MCE.4}
\ee
 The kernel  $K_\textrm{MSM}(\sigma^{\prime}|\sigma)$ has been computed by Kac and Thompson~\cite{KT} obtaining 
\be
K_\textrm{MSM}(\sigma^{\prime}|\sigma) = \delta(\sigma^{\prime} - \sigma), \quad \text{for} \quad \sigma < \sigma_c,
\label{MCE.2}
\ee
and for $\sigma > \sigma_c$
\be
K_\textrm{MSM}(\sigma^{\prime}|\sigma)    = \begin{cases}
     0, \quad $for$ \quad \sigma^{\prime} < \sigma_c,\\
     \phantom{x} & \phantom{x}\\
     \frac{\exp \left \{-\frac{\sigma^{\prime} - \sigma_c}{2(\sigma - \sigma_c)} \right \} }
          {\sqrt{2\pi (\sigma^{\prime} - \sigma_c)(\sigma - \sigma_c)}}, \quad \text{for} \quad
          \sigma^{\prime} > \sigma_c.
\end{cases}
\label{MCE.3}
\ee
Therefore, using the above result together with Eq.~(\ref{ensmbl.001}), one can check that 
for $\sigma < \sigma_c$ Eq.~(\ref{MCE.4}) is solved by
\be
        P_\textrm{SM}({\psi_0}|\sigma)  = \delta(\psi_0), 
                \label{MCE.5}
        \ee
while for $\sigma > \sigma_c$ one gets
\be
        \frac{\exp \left \{-\frac{\psi_0^2}{2(\sigma-\sigma_c)} \right \} }
        {\sqrt{2\pi (\sigma - \sigma_c)} }=
        \int_{\sigma_c}^{\infty} d\sigma^{\prime} \, P_\textrm{SM}({\psi_0}|\sigma^{\prime})\,
\frac{\exp \left \{-\frac{\sigma^{\prime} - \sigma_c}{2(\sigma - \sigma_c)} \right \} }
     {\sqrt{\sigma^{\prime} - \sigma_c}},  
     \label{MCE.6}
     \ee
from which follows 
     \begin{eqnarray}
     P_\textrm{SM}({\psi_0}|\sigma)&=& \sqrt{\sigma-\sigma_c} \; \delta[\sigma-\sigma_c-\psi_0^2]\nonumber \\
     &=&\frac{1}{2}[\delta(\psi_0 + \sqrt{\sigma - \sigma_c})
       + \delta(\psi_0 - \sqrt{\sigma - \sigma_c})]. \nonumber \\
     \end{eqnarray}

\section{ The CE - GCE connection in the ideal Bose gas}
\label{appD}

The Fock space representation of the Hamiltonian $\widehat{{\cal H}} = \sum_{\vec k} \epsilon_k \widehat{a}^{\dagger}_{\vec k}\widehat{a}_{\vec k}$ and number operator 
$\widehat{{\cal N}} = \sum_{\vec k} \widehat{a}^{\dagger}_{\vec k}\widehat{a}_{\vec k}$ is given by
\be
\widehat{{\cal H}} = \sum_{\boldsymbol{n}} E(\boldsymbol{n}) \ketbra{\boldsymbol{n}}{\boldsymbol{n}},
\label{energy.1}
\ee
\be
\widehat{{\cal N}} = \sum_{\boldsymbol{n}} N(\boldsymbol{n}) \ketbra{\boldsymbol{n}}{\boldsymbol{n}},
\label{number.1}
\ee
where $\boldsymbol{n} $ stands for a collection $[n_{\vec k}]$ of occupation numbers, 
$\ket{\boldsymbol{n}}$ are the product states $\prod_{\vec k}\ket{n_{\vec k}}$ and the eigenvalues are given by
\be
E(\boldsymbol{n}) = \sum_{\vec k} \epsilon_k n_{\vec k},
\label{eig.1}
\ee
\be
{\cal N}(\boldsymbol{n}) = \sum_{\vec k}  n_{\vec k}.
\label{eig.2}
\ee
It is first convenient to lay out the general relation between the Fock-space and the Glauber-Sudarshan representations~\cite{Glauber,Sudarshan} of the density matrix
\be
D = \sum_{\boldsymbol{n}} P(\boldsymbol{n})\ketbra{\boldsymbol{n}}{\boldsymbol{n}} = \int d^2\boldsymbol{\alpha} \, P(\boldsymbol{\alpha})
\ketbra{\boldsymbol{\alpha}}{\boldsymbol{\alpha}},
\label{gen.1}
\ee
where $\ket{\boldsymbol{\alpha}} = \prod_{\vec k} \ket{\alpha_{\vec k}}$. The $\vec k$-mode coherent state $\ket{\alpha_{\vec k}}$ is an eigenvector of the annihilation operator $\widehat{a}_{\vec k}\ket{\alpha_{\vec k}}=\alpha_{\vec k} \ket{\alpha_{\vec k}}$ with complex eigenvalue $\alpha_{\vec k}$. The weight functions are related by
\be
P(\boldsymbol{n}) = \int d^2\boldsymbol{\alpha} \, R(\boldsymbol{n}|\boldsymbol{\alpha})P(\boldsymbol{\alpha}),
\label{gen.2}
\ee
with the kernel
\be
R(\boldsymbol{n}|\boldsymbol{\alpha}) = |\braket{\boldsymbol{n}}{\boldsymbol{\alpha}}|^2 = \prod_{\vec k} e^{-|\alpha_{\vec k}|^2} \frac{|\alpha_{\vec k}|^{2n_{\vec k}}}{(n_{\vec k})!}.
\label{gen.3}
\ee

The canonical and the grand canonical ensembles are obtained by 
controlling the density of particles $\rho$ either strictly or on average. Then, the 
corresponding density matrices are given by
\be
D_\textrm{CE}(\rho) = \sum_{\boldsymbol{n}} P_\textrm{CE}(\boldsymbol{n}|\rho)\ketbra{\boldsymbol{n}}{\boldsymbol{n}},
\label{can.1}
\ee
\be
D_\textrm{GCE}(\rho) = \sum_{\boldsymbol{n}} P_\textrm{GCE}(\boldsymbol{n}|\rho)\ketbra{\boldsymbol{n}}{\boldsymbol{n}},
\label{gcan.1}
\ee
where the weight functions read
\be
P_\textrm{CE}(\boldsymbol{n}|\rho) = \frac{1}{Z_\textrm{CE}}e^{-\beta E(\boldsymbol{n})} \, \delta_{{\cal N},V\rho},
\label{can.2}
\ee
\be
P_\textrm{GCE}(\boldsymbol{n}|\rho) = \frac{1}{Z_\textrm{GCE}}e^{-\beta [E(\boldsymbol{n}) - \mu {\cal N}(\boldsymbol{n})]}.
\label{gcan.2}
\ee
In the latter one the chemical potential $\mu$ is fixed by the condition
\be
\frac{1}{V} \langle {\cal N} \rangle_\textrm{GCE} = \rho.
\label{rho.1}
\ee
By going through the same algebra as in the preceding section 
it is straightforward to verify that these weights obey the relation analogous to
Eq.~(\ref{MCE.1})
\be
P_\textrm{GCE}(\boldsymbol{n}|\rho) = \int_0^\infty d \rho^{\prime} \, P_\textrm{CE}(\boldsymbol{n}|\rho^{\prime})
K_\textrm{GCE}(\rho^{\prime}|\rho),
\label{K.1}
\ee
where $K_\textrm{GCE}(\rho^{\prime}|\rho)$ is the probability that the particle density takes the value $\rho^{\prime}$ in the GCE controlled by the average value $\rho$, 
\be
K_\textrm{GCE}(\rho^{\prime}|\rho) = \sum_{\boldsymbol{n}}
P_\textrm{GCE}(\boldsymbol{n}|\rho) \, \delta_{{\cal N}(\boldsymbol{n}),V\rho^{\prime}}.
\label{rel.1}
\ee
This is given by the Kac function~\cite{Ziff}, which for $\rho < \rho_c$ reads
\be
K_\textrm{GCE}(\rho^{\prime}|\rho) = \delta(\rho^{\prime} - \rho), \quad \text{for} \quad \rho < \rho_c,
\label{K.2}
\ee
while for $\rho > \rho_c$ is given by
\be
K_\textrm{GCE}(\rho^{\prime}|\rho)    = \begin{cases}
     0, \quad $for$ \quad \rho^{\prime} < \rho_c,\\
     \phantom{x} & \phantom{x}\\
     \frac{\exp \left \{-\frac{\rho^{\prime} - \rho_c}{\rho - \rho_c} \right \}}
          {\rho - \rho_c}, \quad \text{for} \quad
          \rho^{\prime} > \rho_c.
\end{cases}
\label{K.3}
\ee

Next, inserting Eq.~(\ref{gen.2}) into Eq.~(\ref{K.1}) and taking into account that
the kernel  $R(\boldsymbol{n}|\boldsymbol{\alpha})$ is positive definite, the analogous relation is found to hold between the P-representation weight functions 
\be
P_\textrm{GCE}(\boldsymbol{\alpha}|\rho) = \int_0^\infty d \rho^{\prime} \, P_\textrm{CE}(\boldsymbol{\alpha}|\rho^{\prime})
K_\textrm{GCE}(\rho^{\prime}|\rho).
\label{K.4}
\ee
Integrating this over all $\alpha_{\vec k \neq 0}$, eventually one finds
\be
P_\textrm{GCE}(\alpha_0|\rho) = \int_0^\infty d \rho^{\prime} \, P_\textrm{CE}(\alpha_0|\rho^{\prime})
K_\textrm{GCE}(\rho^{\prime}|\rho),
\label{K.5}
\ee
where the left hand side term is given by~\cite{Glauber}
\be
P_\textrm{GCE}(\alpha_0|\rho) = \frac{1}{\pi \langle n_0 \rangle} e^{-\frac{|\alpha_0|^2}{\langle n_0 \rangle}},
\label{K.6}
\ee
with
\be
\langle n_0 \rangle = [e^{-\beta \widetilde{\mu}(\rho)}-1]^{-1}    = \begin{cases}
     O(1), \quad $for$ \quad \rho < \rho_c,\\
     \phantom{x} & \phantom{x}\\
     V(\rho - \rho_c), \quad \text{for} \quad
          \rho > \rho_c,
\end{cases}
\label{K.7}
\ee
having denoted by $\widetilde{\mu}(\rho)$ the solution of Eq.~(\ref{rho.1}) with respect to $\mu$.
Therefore, defining $\eta_0 = \alpha_0/\sqrt{V}$ and inserting Eqs. (\ref{K.2},\ref{K.3},\ref{K.5},\ref{K.6}) into Eq. (\ref{K.4}), it is easy to verify that in the large $V$ limit for $\rho < \rho_c$
\be
P_\textrm{GCE}(|\eta_0||\rho) = P_\textrm{CE}(|\eta_0||\rho) = \delta(|\eta_0|),
\label{K.8}
\ee
while for $\rho > \rho_c$
\be
P_\textrm{GCE}(|\eta_0||\rho) = \frac{1}{\pi(\rho - \rho_c)} e^{-\frac{|\eta_0|^2}{\rho - \rho_c}},
\label{K.9}
\ee
and
\be
P_\textrm{CE}(|\eta_0||\rho) = \frac{1}{\pi} \delta(|\eta_0|^2 - (\rho - \rho_c)).
\label{K.10}
\ee

\end{document}